\documentclass[smallextended]{svjour3}     

\usepackage{graphicx}
\usepackage{mathptmx}      
\usepackage[english] {babel}
\usepackage[utf8] {inputenc}
\usepackage[T1]{fontenc}
\usepackage{graphicx}
\usepackage{amssymb,amsmath}
\usepackage{color}

\input xy
\xyoption{all}

\providecommand{\abs}[1]{\lvert#1\rvert}

\providecommand{\braket}[3]{\langle #1\mid #2\mid #3  \rangle}
\providecommand{\tensor}{ \ensuremath{ \otimes } }
\providecommand{\overprod}[1]{ \ensuremath{ \odot_{#1} } }

\providecommand{\bM}[1]{#1^{\Box}}
\providecommand{\dM}[1]{#1^{\diamond}}

\journalname{Journal of Statistical Physics}
\begin{document}

\title{Renormalization of cellular automata and self-similarity
}


\author{E. Edlund \and M. Nilsson Jacobi}
\authorrunning{E. Edlund \etal}


\institute{         
E. Edlund \email{erik.edlund@chalmers.se} 
  \and
  M. Nilsson Jacobi  \email{mjacobi@chalmers.se}  \at           
  Complex Systems Group, Department of  Energy and Environment, Chalmers University of Technology, SE-41296 G\"oteborg, Sweden
}

\date{Received: date / Accepted: date}

\maketitle

\begin{abstract}
We study self-similarity in one-dimensional probabilistic cellular automata (PCA) using the renormalization technique. We introduce a general framework for algebraic construction of renormalization groups (RG) on cellular automata and apply it to exhaustively search the rule space for automata displaying dynamic criticality. 

Previous studies have shown that there exists several exactly renormalizable deterministic automata. We show that the RG fixed points for such self-similar CA are unstable in all directions under renormalization. This implies that the large scale structure of self-similar deterministic elementary cellular automata is destroyed by any finite error probability. 

As a second result we show that the only non-trivial critical PCA are the different versions of the well-studied phenomenon of directed percolation. We discuss how the second result supports a conjecture regarding the universality class for dynamic criticality defined by directed percolation.

\end{abstract}

\section{Introduction}
\label{intro}

Elementary PCA, i.e. (1+1)-dimensional, nearest neighbor two-state probabilistic cellular automata, are often used to model an interesting class of critical dynamic behavior referred to as directed percolation. It has been conjectured \cite{hinrichsen_non-equilibrium_2006} that any 
process which
\begin{itemize}
\item has a continuous phase transition from a fluctuating phase into a unique absorbing state;
\item is characterized by a positive one-component order parameter near this transition;
\item has only short range interactions and
\item has no unconventional attributes such as additional symmetries or quenched randomness
\end{itemize}
belong to the directed percolation universality class. In this paper we introduce a general renormalization framework to explore if there are other PCA that show critical behavior, which would contradict the stated conjecture.

In a recent study, Israeli and Goldenfeld \cite{israeli_coarse-graining_2006} explored possible coarse-graining of elementary deterministic cellular automata (CA). The coarse-graining procedure consisted of projections of the state space where configurations in local neighborhoods of states was mapped onto a coarse-grained neighborhood, or to be more precise a coarse-grained light cone. It was shown that  many of the elementary cellular automata can be mapped onto each other by this procedure. Especially  $21$ of the $256$ elementary CA are self-similar and can be mapped onto themselves by an appropriate projection. It is interesting to ask in which sense the dynamics of these CA can be viewed as critical. In this paper we address this question using real-space renormalization of PCA, which includes the usual CA as deterministic limits. 
 
\section{Renormalization of probabilistic cellular automata}
\label{Renorm}


We first formulate the dynamics of cellular automata in terms of matrices. We then present an algebraic framework for renormalization of probabilistic cellular automata on square and diamond lattices. 


\subsection{Matrix formulation of cellular automata}
\label{Mformulation}

Cellular automata (CA) are discrete deterministic dynamical systems with local interactions, i.e. a collection of cells $\{\sigma_i(t)\}$ situated on a lattice together with a local update rule. Each cell can be in one of a finite number of states $\{0,1,...,S-1\}$. The update rule is applied synchronously and independently on every neighborhood of cells. The concept is easily generalized to probabilistic cellular automata (PCA) by considering non-deterministic update rules.

For our purposes, it is advantageous to use a spin representation and consider a matrix form of the update rule. Each local state, or spin, is represented by a vector~$s$ of length $S$ corresponding to a column of the identity matrix in $\mathbf{R}^S$. Collections of spins are represented by tensor products of their constituents, $\{\sigma_1,\sigma_2,...\} \leftrightarrow \left(s_1,s_2,... \right) = s_1 \tensor s_2 \tensor \cdots$. 

The size of the neighborhood influencing each spin depends on the dimension and topology of the lattice. Let the number of such spins be denoted $z$. We then define the matrix $P$ as the $S\times S^z$ matrix which corresponds to the local update rule for a single cell. For deterministic cellular automata, each entry of $P$ is 0 or 1, but in the PCA generalization $P$ is a probability matrix ($P_{ij}\in [0,1]$ and $\sum_i P_{ij} = 1$). An update of a spin is achieved by the matrix product where $P$ operates on the influencing neighborhood, $s_i(t+1) = P\cdot \prod_{\langle i,j \rangle} s_j(t)$ where the product should be interpreted in terms of tensor products and $\langle i , j \rangle$ denotes neighboring cells.

As an example, consider Wolfram's elementary cellular automata \cite{wolfram_statistical_1983}. They have two states which we represent with $(1,0)$ if $\sigma=1$ and with $(0,1)$ if $\sigma=0$. Further, they are defined on a square lattice so every spin $s_i$ has three influencing preceding neighbors $(s'_{i-1},s'_{i},s'_{i+1})$. This neighborhood is a vector of length $2^3$ and $P$ is a $2\times2^3$ matrix. As $P$ is a probability matrix, its first row has eight parameters $P_1 = (p_1,p_2,...p_8) \in [0,1]^8$ and its second consists of the corresponding $1-p_i$. For example, $P_1= (0,1,0,1,1,0,1,0)$ corresponds to rule $90$ in Wolfram's nomenclature \cite{wolfram_statistical_1983} as can be seen by successive application to the different spin configurations.

 To coarse-grain an automaton updates of larger neighborhoods must be considered. These aggregated update rules also have matrix representations, possible to express in terms of $P$. Let the matrix that updates $n$ neighboring spins from their combined neighborhood be denoted by $P_n$. As the update rule is applied independently and in parallel over the lattice, $P_n$ can be constructed in terms of a kind of tensor product of $P$:s. However, the neighborhoods will overlap so the usual tensor product cannot be used, easily seen e.g.\ by considering the dimensions of $P\tensor P \tensor \cdots P$.

Consider two adjacent spins on a square lattice. They have three influencing neighbors each in their nearest neighbor regions. Only four of these are unique as the neighborhoods overlap by two spins. Define a 2-overlapping tensor product for $2\times2^3$ spin matrices $A$ and $B$ as
\begin{equation}
\braket{s_1',s_2'}{A\overprod{2} B}{s_0,s_1,s_2,s_3} = 
\braket{s_1'}{A}{s_0,s_1,s_2} \cdot \braket{s_2'}{B}{s_1,s_2,s_3}
\end{equation}
 in Dirac's vector notation \cite{sakurai_modern_1993}. With this notation $P_2$ can be expressed in terms of the basic update matrix as $P\overprod{2}P$.

More generally, we define a $k$-overlapping tensor product for spin matrices $A$ and $B$ as 
\begin{eqnarray}
\nonumber&\braket{s_1',...,s_{n_1+m_1}'}{A\overprod{k} B}{s_0,s_1,...,s_{n_2+m_2-k}} =\\ 
\label{overprod}
&\braket{s_1',...,s_{n_1}'}{A}{s_0, ... ,s_{n_2}} \cdot \braket{s_{n_1+1}',...,s_{n_1+m_1}'}{B}{s_{n_2-k},...,s_{n_2+m_2-k}}
\end{eqnarray}

where $A$ ($B$) describe the transition of $n_2$ ($m_2$) spins into $n_1$ ($m_1$) ones. The matrix representation of the update rule for $n$ adjacent spins, $P_n$, on a lattice where adjacent spins have $k$ overlapping neighbors is now given by a product of $n$ $P$:s, $P_n = P\overprod{k}P\overprod{k}\cdots P$.


\subsection{Coarse-graining of probabilistic cellular automata}
\label{RGPCA}

The matrix representation can be used to calculate both coarse-graining and renormalization of probabilistic cellular automata on square and diamond lattices. We denote the matrices defining the dynamics by $\bM{P}$ and $\dM{P}$ respectively. The coarse-graining transformation consists of a projection of a block of $N$ cells into a single cell in space as well as a stroboscopic coarse-graining in time. This combination is needed to keep the structure of the light cone. The projection can be written as a $S\times S^N$ matrix $\Pi$. The goal is to calculate the effective dynamics on the coarse-grained level, which also has an $S \times S^z$  matrix representation $\tilde{P}$. 

The transformation has two steps. First we define a cellular automaton on blocks of cells on the original lattice. This CA will have an alphabet of size $S^N$ and each time step will correspond to $N$ time-steps of the original CA. We denote the dynamics on the block level by  $\bM{Q}$ on the square lattice and by $\dM{Q}$ on the diamond lattice. This first step is straightforward and there are no formal requirements on the dynamics. In the second step, the alphabet over blocks of states is projected onto the original one-block alphabet, which results in a coarse-grained version on the same form as the original CA. For the second step to result in a well defined dynamics, i.e.\ for the state of the automaton at time $t+1$ to be independent of the state at time $t-1$ given the state at time $t$ (the Markov property), there are restrictions on both the projection and the dynamics \cite{israeli_coarse-graining_2006,jacobi_spectral_2009}. These restrictions define the coarse-graining and can be formulated in terms of the matrices $\Pi$ and $P$. Later when we discuss renormalization the coarse-graining restrictions are only approximately fulfilled.

\begin{figure}
\begin{center}
\includegraphics[width=\linewidth]{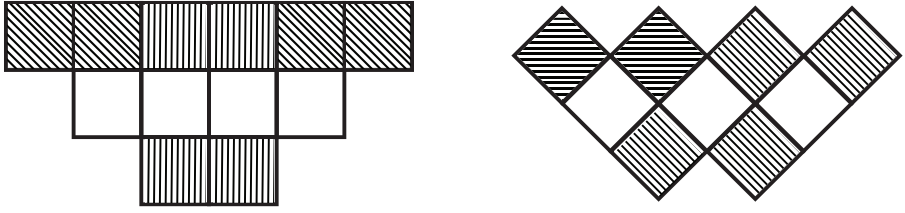}
\caption{Left: light cones, defined as the influencing neighborhoods of a given cell, on a square and a diamond lattice. Right: blocks used for a coarse-graining of a PCA on a square lattice. }
\label{lightcones}
\end{center}
\end{figure}

Due to the $45^{\circ}$ angle of the light cones on the lattices, $n$ adjacent spins will have $n+1$ influencing neighbors on a diamond lattice and $n+2$ on a square lattice. For a block size of $N$, the block CA corresponds to $N$ time steps from $3N$ onto $N$ spins on the original lattice for the square case. Figure \ref{lightcones} illustrates the case $N=2$. 
On a diamond lattice the influencing neighborhood consists of $2N$ spins. The transition matrix for the block CA, denoted by $Q$, can be concisely expressed in terms of the overlapping tensor product of equation \eqref{overprod}. On the diamond lattice we have 
\begin{equation}
\label{dQ}
  \dM{Q} = \dM{P}_N \cdot \dM{P}_{N+1} \cdots \dM{P}_{2N-1}
\end{equation}
and on the square lattice $\bM{Q} = \bM{P}_N \cdot \bM{P}_{N+2} \cdots \bM{P}_{3N-2}$. The description of $\bM{P}_N$ in terms of $\bM{P}$ was given in the previous section as an tensor product with overlap $2$. On the diamond lattice the overlap of neighborhoods of adjacent spins is $1$, so the corresponding $\dM{P}_N $ are given by $\dM{P}_N = \dM{P} \overprod{1}  \dM{P} \overprod{1}  \cdots \dM{P}$.

\begin{eqnarray}
\label{commuting_diagram}
\xymatrix{
S \times S  \ar[rr] ^{\dM{\widetilde{P}} } & & S \\
S^N \times S^N \ar[rr] ^{\dM{Q}} \ar[u] ^{\Pi \otimes \Pi } & & S ^N \ar[u]  ^{\Pi} }
\end{eqnarray}

For a projection $\Pi$ to form a well defined coarse-grained dynamics, the diagram (\ref{commuting_diagram}) must commute in the sense that a time evolution through $Q$ followed by a projection should give the same result as a projection followed by a time evolution through $\widetilde{P}$, see \cite{jacobi_spectral_2009} for details.  Algebraically this means that there must exist an $S \times S^2$  matrix $\dM{\widetilde{P}}$, representing the coarse-grained dynamics, such that 
\begin{equation}
\label{projectBmatrix}
\Pi \cdot \dM{Q}  =\dM{\widetilde {P}} \cdot ( \Pi \tensor \Pi)
\end{equation}
and $\Pi \cdot \bM{Q}  =\bM{\widetilde {P}} \cdot (\Pi \tensor \Pi \tensor \Pi )$. If $\dM{\widetilde{P}}$ exists, then equation~\ref{projectBmatrix} can be solved using a pseudoinverse\footnote{The pseudoinverse used is the Moore-Penrose pseudoinverse for rectangular matrices, i.e. for a $n\times m$ matrix $A$ with $m<n$: $A^+ = (A^T A)^{-1} A^T$.}  $( \Pi \tensor \Pi ) ^+$,  
\begin{equation}
\label{projectD}
\dM{\widetilde {P}} = \Pi \cdot \dM{Q}  \cdot (\Pi \tensor \Pi)^+ ,
\end{equation}
and correspondingly $\bM{\widetilde {P}}  = \Pi \cdot \bM{Q}  \cdot (\Pi \tensor \Pi \tensor \Pi) ^+$ on the square lattice. 

\subsection{Renormalization of probabilistic cellular automata}

When applying renormalization theory one is usually much less concerned with using exact coarse-grainings. As long as a projection which preserves the symmetries of the problem is used one usually gets correct quantitative behavior. With this in mind it is natural to use equation \eqref{projectD} as an approximate effective dynamics, ignoring that equation~\eqref{projectBmatrix} is usually not fulfilled. The renormalization group transformation $\mathcal{R}_N$ is then given by equations~\eqref{dQ} and~\eqref{projectD} as 
\begin{equation}
\label{RGd}
\mathcal{R}_N(\dM{P}) \equiv \Pi \cdot  \dM{P}_N \cdot \dM{P}_{N+1} \cdots \dM{P}_{2N-1} \cdot (\Pi \tensor \Pi)^+
\end{equation}
on the diamond lattice and
\begin{equation}
\label{RGb}
\mathcal{R}_N(\bM{P}) \equiv  \Pi \cdot \bM{P}_N \cdot \bM{P}_{N+2} \cdots \bM{P}_{3N-2}  \cdot (\Pi \tensor \Pi \tensor \Pi) ^+
\end{equation}
on the square lattice.

The pseudo-inverse can be interpreted as a back-projection from the coarse-grained description to the corresponding states in the original automaton. If diagram \ref{commuting_diagram} commutes then any such state will result in the same subsequent dynamics. If it does not then the way in which the back-projection is performed affects the results of the renormalization.  The back-projection thus should be done in a way which is consistent with the dynamics of the original automaton. This is achieved by weighing with the stationary distribution over different states.

Constructing a diagonal matrix $\bM{D}_{eq}$ with the stationary distribution over $3N-2$ spins on the diagonal gives us the weighed renormalization transformation on the square lattice
\begin{equation}
\label{RGbw}
\mathcal{R}_N(P) = 
\Pi \cdot \bM{P}_N \cdot \bM{P}_{N+2} \cdots \bM{P}_{3N-2}  \cdot (\Pi \tensor \Pi \tensor \Pi \cdot \bM{D}_{eq}) ^+
\end{equation}
On the diamond lattice the stationary distribution is taken over $2N-1$ spins and the resulting equation is 
\begin{equation}
\label{RGdw}
\mathcal{R}_N(P) = 
\Pi \cdot \dM{P}_N \cdot \dM{P}_{N+1} \cdots \dM{P}_{2N-1}  \cdot (\Pi \tensor \Pi \cdot \dM{D}_{eq}) ^+
\end{equation}

Denote the stationary distribution over $M$ adjacent spins by $s^M_{eq}$. Since the stationary distribution must be invariant under the time evolution, it must satisfy the equation
\begin{equation}
\label{eq.distr.}
s^M_{eq} = P_M \cdot s^{M+l}_{eq}
\end{equation}
with $l=1$ on the diamond and $l=2$ on the square lattice. This is not a closed equation as determination of the stationary distribution for $M$ spins require knowledge of the distribution for $M+l$ spins. The solution lies in approximating the latter in terms of the former. This amounts to disregarding statistical correlations over neighborhoods larger than $M$ spins. The natural approximation\footnote{It can be shown that this is the maximum entropy distribution over $M+1$ spins such that it reduces to the distribution over $M$ spins when summed over the first or last spin.} is given by
\begin{equation}
\label{BBGKYd}
(s_1,s_2,...,s_{M+1})_{eq} = 
\frac{(s_1,...,s_M)_{eq}\cdot(s_2,...,s_{M+1})_{eq}}{(s_2,...,s_M)_{eq}}
\end{equation}
on the diamond lattice and
\begin{equation}
\label{BBGKYb}
(s_1,s_2,...,s_{M+2})_{eq} =
\frac{(s_1,...,s_M)_{eq}\cdot(s_2,...,s_{M+1})_{eq}\cdot(s_3,...,s_{M+2})_{eq}}{(s_2,...,s_M)_{eq}\cdot(s_3,...,s_{M+1})_{eq}}
\end{equation}
on the square lattice. To calculate the stationary distribution of for example $3N-2$ spins to order $M$ we solve equation~\eqref{eq.distr.} using equation~\eqref{BBGKYb} and then sum over spins to arrive at $s^{3N-2}_{eq}$. This gives increasingly better approximations for larger $M$, taking larger statistical correlations into account. This defines a hierarchy of approximations, where the unweighed inverse projections of equations~\eqref{RGd} and \eqref{RGb} can be seen as zeroth order. 


We conclude this section by recalling some of the key points of renormalization theory. An introduction to the subject is found in for example  \cite{goldenfeld_lecturesphase_1992}. As macroscopic observations of a system correspond to  a severely coarse-grained version of the microscopic dynamics the fixed points of the transformation are central. For the cellular automata these are the ones described by a matrix  $P^*$ such that 
\begin{equation}
\mathcal{R}_N(P^*) = P^*
\end{equation}

By repeated application of the renormalization transformation, each reducing length scales of the system, it is easy to see that for example the correlation length at a fixed point is either zero or infinite. The latter corresponds to a system at criticality with self-similarity at all scales. The former corresponds to bulk phases of the system where the dynamics are either pure noise, corresponding to a high-temperature limit, or frozen, corresponding to a low-temperature limit. These are known as trivial fixed points.

For systems close to a fixed point the behavior is governed by the eigenvalues of the Jacobian of the renormalization transformation as can be seen by linearizing around the fixed point,
\begin{equation}
\widetilde{P} = P^* + J(P^*)\cdot \delta P + O((\delta P)^2)
\end{equation}
where $ J(P^*)$ is the Jacobian of $\mathcal{R}_N$. Components of $\delta P$ which lie along eigenvectors of $J$ with corresponding eigenvalues $\abs{\lambda} < 1$ will shrink. Such directions are called \emph{irrelevant} as they eventually disappear after repeated applications of $\mathcal{R}_N$. In the same way, directions along which the eigenvalues of $J$ have magnitude larger than one are called \emph{relevant} as such deviations grow exponentially and drive $P$ away from $P^*$. The relevant degrees of freedom are associated with control parameters that must be tuned to achieve critical behavior.

The universality observed among widely disparate systems close to criticality stems from the fact that it is the same eigenvalues (i.e. the same control parameters) which drives the systems near the fixed point, regardless of their respective microscopic dynamics. These eigenvalues determine the critical exponents of systems near their critical point.


\subsection{Searching for renormalizable models}
\label{search}

One of the advantages with cellular automata is their discreteness and simplicity. It allows us to explore all possible (non-stochastic) renormalization projections for a given block size. We can search the space of PCA for fixed points of each such projection and in principle exhaustively enumerate all renormalizable cellular automata in a given class. This provides a program for determining the possibility of self-similarity in any given class of cellular automata. 

We here restrict our exposition to projections of blocks of size two. This is motivated by the fact that renormalization theory shows that the actual change in length scales does not matter as long as the projection preserves the relevant symmetries of the problem. In this setting this means that the only possible difference between renormalizations of blocks of size two and larger ones lies in possibilities of new symmetry preservations. For example, a block two projection cannot constitute a proper majority rule such as is usually used for renormalization of Ising models. This does not seem to be a problem as shown by the example of compact directed percolation. This universality class shows up even for projections not respecting its $0 \leftrightarrow 1$ symmetry. We have done some investigations of projections of blocks of size three without finding any new results.

For a given probabilistic cellular automaton we want to find fixed points of the corresponding renormalization transformation, $P^*$ such that $P^*=\mathcal{R}_N(P^*)$. Here the advantage of our approach is apparent: using the approximation of equation~\eqref{projectD} and its analog on the square lattice we have polynomial equations in the parameters of $P$. This allows us to quickly search for self-similar automata.

On a diamond lattice with blocks of size two the problem is a system of four polynomial equations in four variables of degree five. We were able to use polynomial solving software to completely solve this system for each of the 14 possible non-trivial projections of two cells into one. We used the hierarchical scheme introduced in section \ref{RGPCA} to refine the approximation for the resulting fixed points which do not fulfill the commuting condition of diagram \ref{commuting_diagram}.

On the square lattice the corresponding system has eight equations of degree eight in as many variables. Our software cannot solve this. Instead we perform local searches for fixed points seeded with random initial conditions for each of the possible projections.

\begin{figure}
\includegraphics[width=\textwidth]{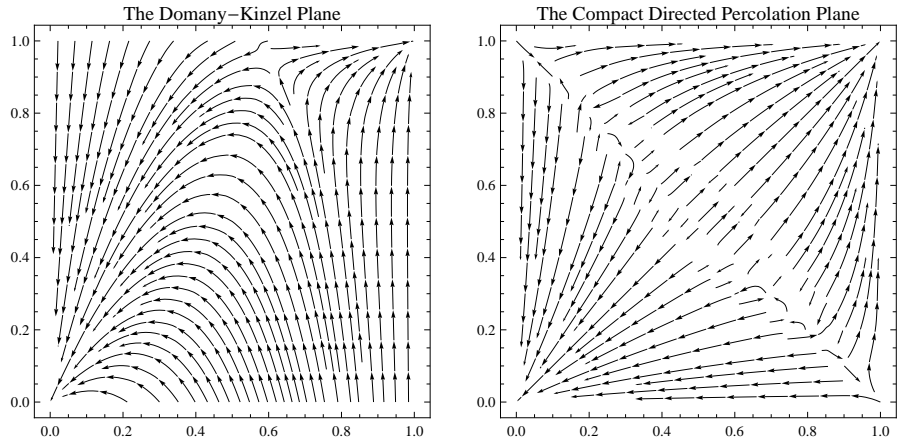}
\caption{\label{flows} The renormalization flow for the automata in the diamond lattice for two planes in the hyper cube. To the left is the Domany-Kinzel~\cite{domany_equivalence_1984} plane, characterized by $p_2=p_3$ and $p_4 = 0$. To the right is the plane characterized by $p_1 = 1- p_4 = 1$, showing the compact directed percolation line.}
\end{figure}

The algebraic method of equation~\eqref{projectD} is fast, but may miss self-similar automata which are fixed points only with the weighing on the  stationary distribution included. Using a fine grid (21 points for each dimension, i.e. $21^4$ total), we compute the vector field induced by equation~\eqref{RGdw} on the diamond lattice and find its zeros for each of the 14 non-trivial projections. Figure~\ref{flows} shows the field for the projection
\begin{equation}
\Pi \,\,=\,\, \left\{ 
\begin{array}{ccl}
(0,0) & \rightarrow & \,0 \\
(0,1), (1,0) (1,1)& \rightarrow & \,1
\end{array}
\right.
\end{equation}
for two planes of the relevant hyper-cube $[0,1]^4$.

\section{Existence and stability of self-similar automata}
\subsection{Overview of results}
\label{results}

Our search shows three kinds of fixed points of the renormalization scheme. They are:
\begin{itemize}
\item Trivial fixed points - e.g. rules in which there is no correlation in time, most of them corresponding to high temperature limits. 
\item Deterministic CA - rules for which all transition probabilities are 0 or 1. These are shown to be unstable and correspond to low temperature limits.
\item Non-deterministic CA - rules with stochastic time evolution, corresponding to finite critical temperatures. All such solutions are shown to belong to the directed percolation (DP) universality class or the compact DP class.
\end{itemize}
As shown in previous work \cite{israeli_coarse-graining_2006}, there is a fairly large number of deterministic automata which coarse-grain themselves. Perhaps a bit surprising, there is no such abundance of non-deterministic ones. In fact, our search finds no non-trivial non-deterministic CA admitting a renormalization with block size 2, apart from models of directed percolation or compact directed percolation.

It is known \cite{hinrichsen_non-equilibrium_2006} that the critical manifold for directed percolation ends in a point at which the universality class changes, called compact directed percolation. The change of universality class is due to an extra $0 \leftrightarrow 1$ symmetry. Our purely algebraic method (equations~\eqref{RGd} and~\eqref{RGb}) cannot find this point, but when weighing is included (equations~\eqref{RGbw} and~\eqref{RGdw}) it is identified correctly. We see that extra symmetry present at this point that changes its universality class is not a problem for the two block projections, which indicate their generality.

\subsection{Trivial fixed points}
In renormalization theory, trivial or weakly coupled fixed points are points where there are no correlations (or trivial correlations in a frozen system).

One example of such a point in our context is an automaton with constant probabilities over all preceding configurations, $P_{ki} = P_{kj} \, \forall i,j$. For such automata the spins behave independently and each take a value $k$ with some probability $P_{k} = p\in [0,1]$. It is not surprising that we find such fixed points for all projections. For projections which assigns more configurations to one state than to another, the trivial fixed points are $p=0$ and  $p=1$, corresponding to rules $0$ and $255$ for the elementary automata. For the majority rules with ties broken in different ways, all $p\in[0,1]$ gives fixed points, while for example the projection

\begin{equation}
\Pi \,\,=\,\, \left\{ 
\begin{array}{rcl}
(0,0), (1,1) & \rightarrow &0 \\
(0,1), (1,0) & \rightarrow &1
\end{array}
\right.
\end{equation}
has fixed points for $p\in\{0,1/2,1\}$.

More generally, any rule where the probabilities only depend on a single spin in the preceding neighborhood is a trivial point as there are no spatial interactions in such an automaton. No non-deterministic automata fitting this description not belonging to the specific case handled above are found as fixed points of any of the renormalizations. However, a number of deterministic automata does; more on those below.

\subsection{Deterministic fixed points}

Coarse-graining of deterministic elementary cellular automata is studied extensively in \cite{israeli_coarse-graining_2006} and also discussed in~\cite{jacobi_spectral_2009}. Among the central results in~\cite{israeli_coarse-graining_2006} is a diagram showing how different such CA coarse-grain each other. Our method reproduce all 21 fixed points of this diagram as expected. We can also analyze these fixed points using renormalization theory.

First of all we see that most self-similar CA are trivial fixed points in the sense described above. The exceptions are rules 60, 90, 102 and their $0\leftrightarrow1$ symmetries as well as rule 150. These rules correspond to critical systems as they build fractal structure at all scales.

\begin{figure}
\includegraphics[width=0.99\textwidth]{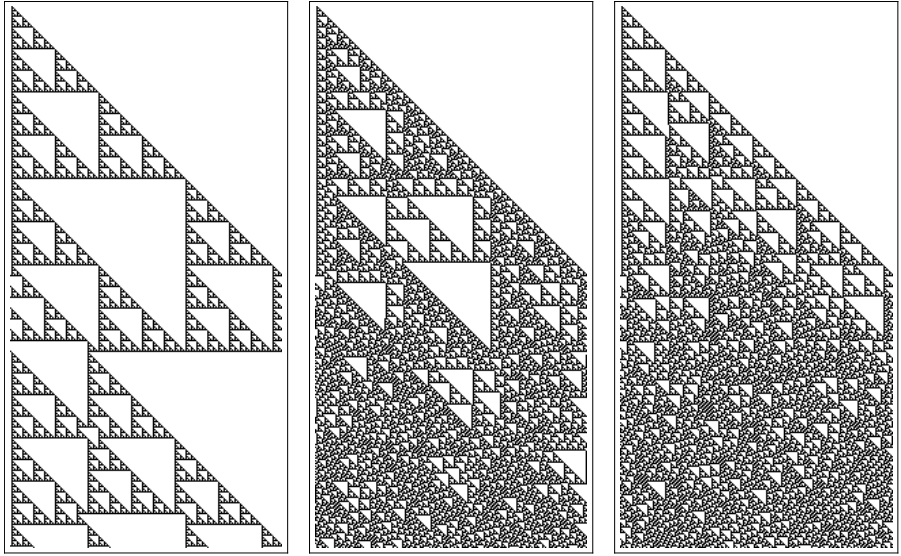}
\caption{\label{DCAnoise}The self-similar structures generated by deterministic automata are destroyed by any finite noise. Time evolution from a single seed for Wolfram's rule 60 without (left) and with perturbations of $0.4\%$ in $p_2$ and $p_3$ (middle and right). Cyclic boundary conditions are used.}
\end{figure}

However, these fixed points are degenerate in another sense. Analyzing the Jacobian of the renormalization transformation at these points, we find that all allowed directions (directions which does not cause the probabilities to leave the interval $[0,1]$) are associated with eigenvalues $\abs{\lambda_i} > 1$. This means that these CA are critical only in the deterministic limit and at large scales the self-similar structure is destroyed by any stochastic component of the dynamics. Figure~\ref{DCAnoise} shows an example of this: the time evolution of rule 60 builds a Sierpinski triangle but a small perturbation to any of the parameters breaks up the large scale structure.

\subsection{Non-deterministic fixed points}

\begin{figure}
\includegraphics[width=0.32\textwidth]{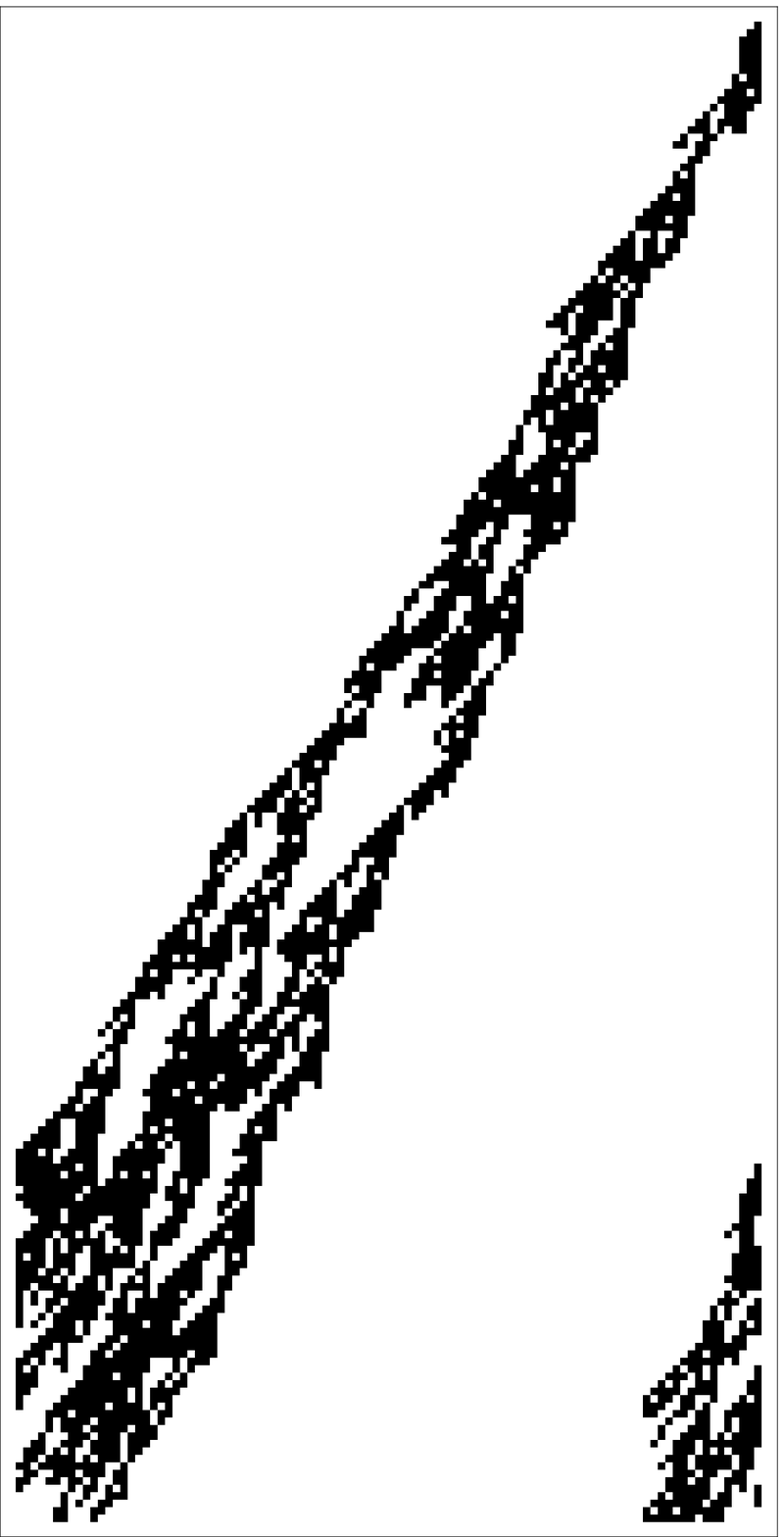}
\includegraphics[width=0.32\textwidth]{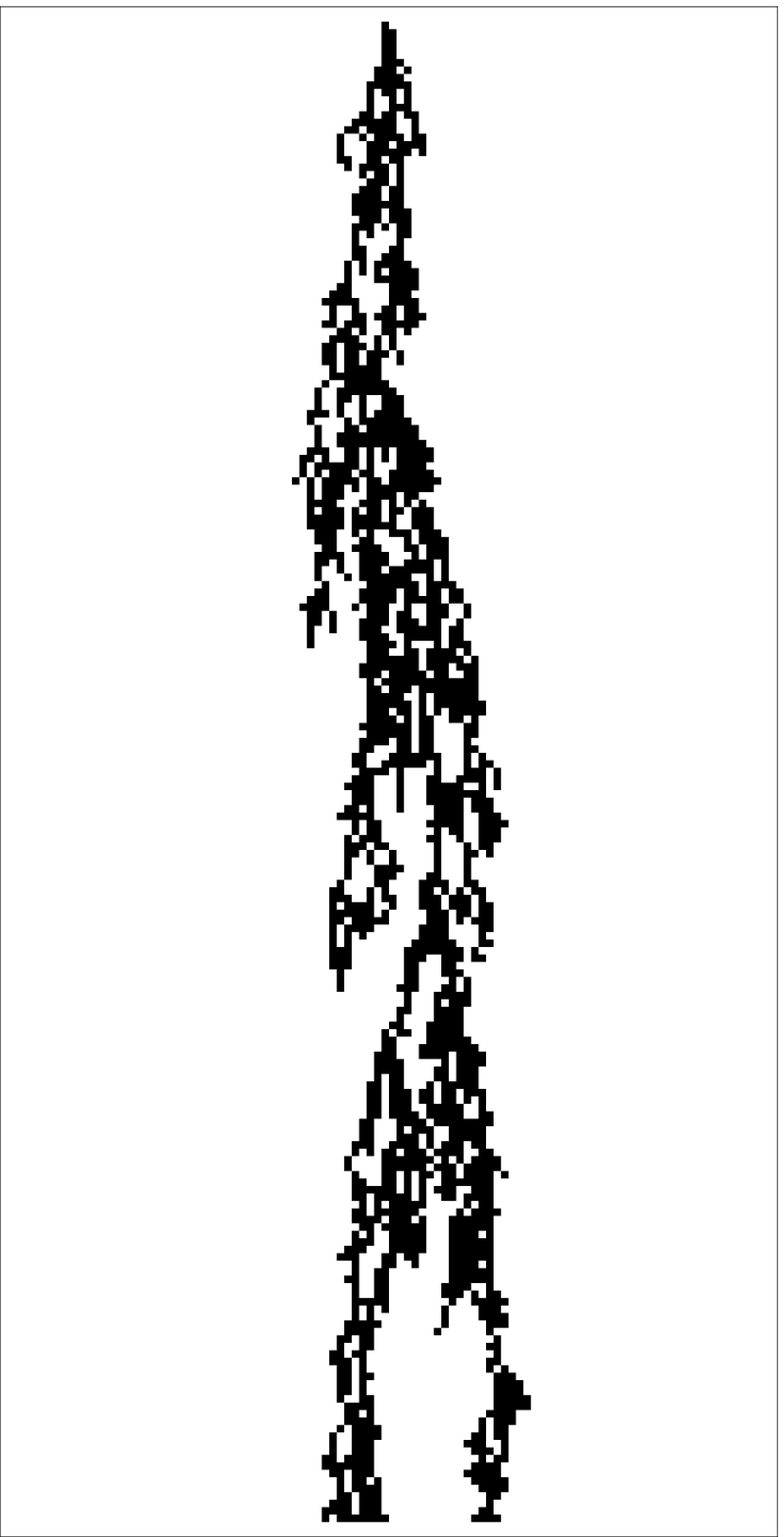}
\includegraphics[width=0.32\textwidth]{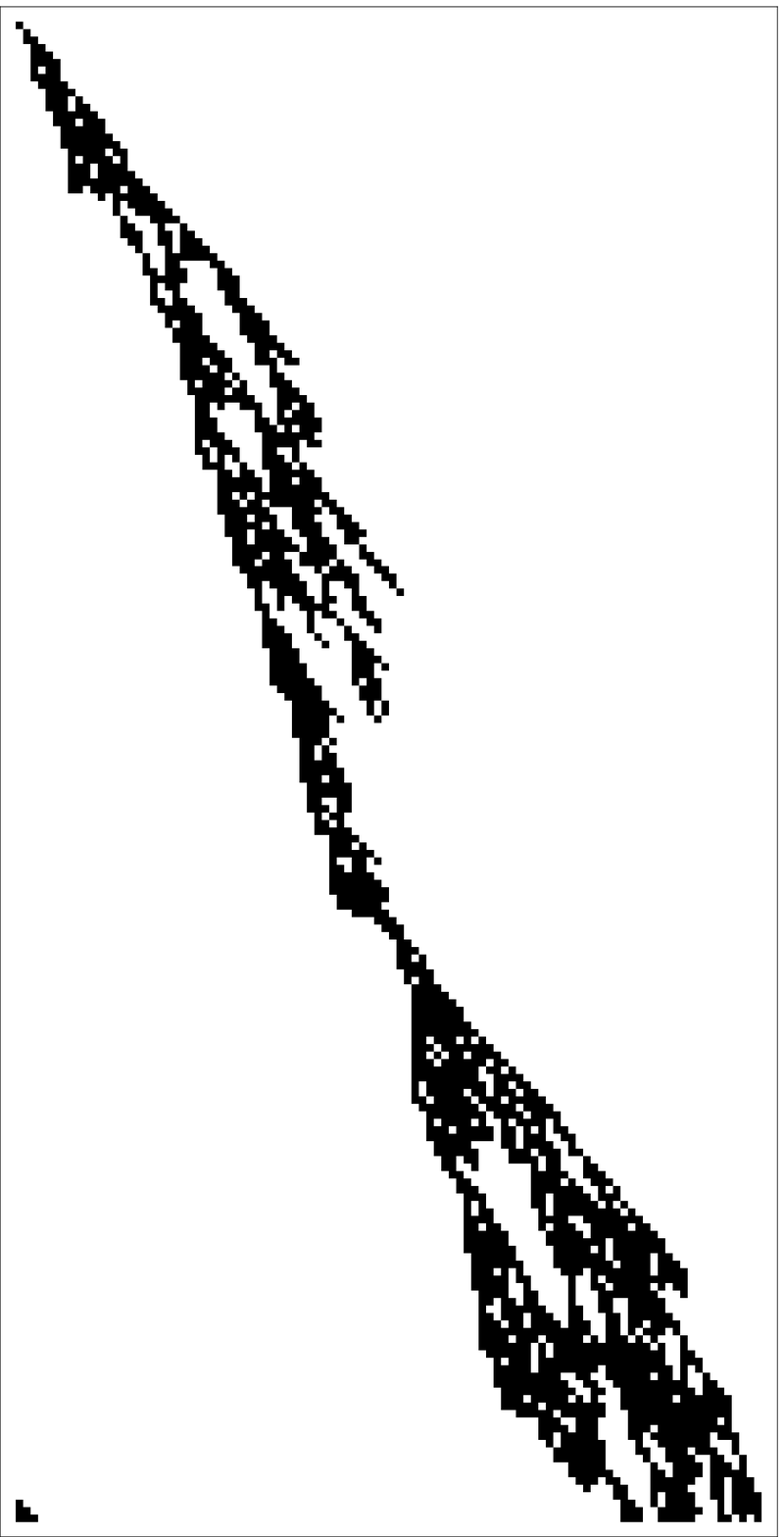}
\caption{\label{DPfig}Time evolution from a single seed for the three different models of directed percolation found as fixed points of the renormalization on a square lattice. The automata shown are slightly above criticality. Cyclic boundary conditions are used.}
\end{figure}

The only non-deterministic fixed points found by our algebraic method are models of directed percolation (DP). On the diamond lattice we find only one, known as the Domany-Kinzel automaton \cite{domany_equivalence_1984}. This point exists in three versions on the square lattice. Figure \ref{DPfig} shows time evolutions of these automata slightly above criticality started from single seeds.

The fixed point on the diamond lattice of the unweighed renormalization of equation \eqref{projectD} is given by $\dM{P}_{1} \approx \left( 0.8389, 0.6096, 0.6096, 0 \right)$. This is for the projection 
\begin{equation}
\Pi \,\,=\,\, \left\{ 
\begin{array}{ccl}
(0,0), (0,1), (1,0)& \rightarrow &0 \\
(1,1) & \rightarrow &1
\end{array}
\right.
\end{equation}
The projection can be understood as one that preserves the absorbing state of the percolation process. Running the automaton shows that this zeroth order approximation underestimates the critical point. We therefore refine our result by using the hierarchy described in section \ref{RGPCA}. Iterating equation \eqref{eq.distr.} combined with equation \eqref{BBGKYd} until convergence and weighing the inverse projection with this distribution gives an implicit equation for a fixed point. We regard the fixed point condition as an optimality criterion and use a simple direct search method, compass search~\cite{kolda_optimization_2003}, to minimize the error. Truncating the hierarchy at $M = 6$ we get the value for the critical point $\dM{P}_{1} \approx \left( 0.9447, 0.5923, 0.5923, 0 \right)$, which turns out to be in better agreement with the correct one as is confirmed by simulations.\footnote{There are several methods to estimate the critical point for numerical directed percolation models. We plot log-log diagrams of the occupancy at time $t$. For a description and discussion, see \cite{hinrichsen_non-equilibrium_2000}.}

We note that while the percolation of the Domany-Kinzel automaton has been studied using a renormalization scheme similar to ours in e.g.~\cite{tom_renormalization_1997}, the larger generality in our scheme allows us to see explicitly that the automaton is only a fixed point if there exists a completely absorbing state, as shown both by the roots of the polynomials and by an analysis of the eigenvectors at the fixed point.

On the square lattice there are three directed percolation fixed points, shown in figure \ref{DPfig}. The non-symmetric fixed points correspond exactly to the Domany-Kinzel automaton if the square lattice is tilted $45^{\circ}$ to give a diamond lattice. In other words, the rules are independent of either the left or right preceding neighbor, 
\begin{equation}
\bM{P}_{1} \approx \left( 0.8389, 0.6096, 0.6096, 0, 0.8389, 0.6096, 0.6096, 0 \right)
\end{equation}
 for the right skewed automaton in the zeroth approximation and similarly for the left skewed one. The corresponding refined approximations are given by the values given above for the diamond lattice. The symmetric fixed point is situated at
 \begin{equation}
\bM{P}_{1} \approx \left( 0.8073, 0.7679, 0.3957, 0.2227, 0.7679, 0.7206, 0.2227, 0 \right)
\end{equation}
and a refinement with a hierarchy truncated at $M=6$ gives the point
 \begin{equation}
\bM{P}_{1} \approx \left( 0.9807, 0.9522, 0.1623, 0.0711, 0.9522, 0.9267, 0.0711, 0 \right)
\end{equation}
We note that the zeroth approximation also has a set of spurious fixed points above the critical point. These disappear when higher order approximations are used.

When we include the weighing systematically in the search for the diamond lattice we find a line of fixed points belonging to the compact directed percolation class. These automata are on the form
 \begin{equation}
 \label{CDP}
\dM{P}_{1} = \left(1, p, 1-p, 0 \right)
\end{equation}
for any $p\in[0,1]$. These possess an extra $0\leftrightarrow1$ symmetry compared to the directed percolation processes, placing them in another universality class. For $p=0.5$ the automaton describes a annihilating random walk with the boundaries interpreted as particles. Tuning the parameter biases the walk in the corresponding direction until, at $p=1$ or $0$, the deterministic limit is achieved and the particles move in straight lines.

Studying the eigenvalues of the transformation around the compact directed percolation points shows that they have no irrelevant directions. They have an exactly marginally relevant direction along the line defined by equation~\eqref{CDP} as any movement along this line will result in a new fixed point being visited. 

The aim of the method presented in this paper is not to calculate the critical exponents for any particular universality class. For completeness we nevertheless examine the possibilities of such a calculation for the directed percolation fixed point. We know that the spatial correlation length scales algebraically close to the critical point as $\xi \sim t^{-\nu_\perp}$. Through renormalization theory we know that $\nu_\perp = \ln{b}/\ln{\lambda}$ where $\lambda$ is the largest eigenvalue of the Jacobian of the renormalization transformation and $b$ is the length scaling (equal to the block size $N$ in our case) \cite{goldenfeld_lecturesphase_1992}.

For the zeroth order approximation we can do the differentiation analytically. For higher order approximations we calculate the Jacobian numerically. Table~\ref{exponent} shows the obtained values for different orders of the approximation when applied to the Domany-Kinzel automaton at the point reported above. They should be compared with the value $\nu_\perp = 1.096854(4)$ reported in~\cite{hinrichsen_non-equilibrium_2006}. Note that already using neighborhoods of size four comes within $10\%$ of the correct value. However, the convergence for larger neighborhoods is extremely slow. This is most likely due to the geometric scaling of the correlations close to criticality. This makes the method unsuitable for accurate determination of the exponents.
\begin{table}
\caption{The largest eigenvalue $\lambda_1$ and corresponding critical exponent $\nu_\perp$ for the renormalization transformation with different orders of approximation $M$.}
\label{exponent} 
\begin{tabular}{lll}
\hline\noalign{\smallskip}
M & $\lambda_1$ & $\nu_\perp$  \\
\noalign{\smallskip}\hline\noalign{\smallskip}
0 & 1.297 & 2.667 \\
4 & 1.994 & 1.004 \\
6 & 1.986 & 1.011 \\
7 & 1.984 & 1.011 \\
8 & 1.984 & 1.012 \\
9 & 1.983 & 1.012 \\
\noalign{\smallskip}\hline
\end{tabular}
\end{table}
Using the same procedure for the fixed points on the square lattice gives qualitatively similar results but with even slower convergence. This is not surprising since the model in this case is embedded in a large space of confounding variables.


\section{Summary and discussion}
\label{discussion}



In this paper we use renormalization theory to show that the deterministic self-similar elementary cellular automata correspond to low temperature limits where any finite probability of errors destroys correlations at large enough scale. This is analogous to the situation for the one-dimensional Ising model which lacks large scale structure for all $T>0$. The situation is quite different in for example the two dimensional Ising model where the critical point is defined by only two relevant parameters, temperature and the external field. Large scale structures are insensitive to other perturbations of the microscopic dynamics, corresponding to irrelevant directions of the renormalization transformation. This is why the Ising model defines a universality class and can reproduce the scaling behavior of a large variety of experimental systems near criticality. We conclude that the self-similar deterministic CA do not define any interesting universality classes and that any perturbation of their dynamics destroys large scale fractal structure. 
Our analysis shows that there are limits on large scale (space-time) structure generated by elementary cellular automata with finite error probability. Lindgren~\cite{lindgren_correlations_1987} showed that the space correlations of automata is destroyed by uniform noise. Our method generalizes this to space-time correlations and noise in any part of the rule of the automaton.

By systematically exploring both the space of probabilistic cellular automata and renormalization transformations we show that the only non-trivial universality class among two-state, one-dimensional nearest-neighbor automata is directed percolation. This results can be seen as further evidence for the directed percolation conjecture discussed in the introduction, i.e. that any critical system describable by two-state nearest neighbor PCA belongs to the same universality class as DP. Our results provide no definite proof of the conjecture as only deterministic projections with a relatively limited range have been considered.

\begin{acknowledgements}
The authors thank Kristian Lindgren for discussions and especially for clarifying several important issues in relation to the transfer matrix formalism.
\end{acknowledgements}

\bibliographystyle{spmpsci}      
\bibliography{renormBibl}

\end{document}